\documentclass{aastex}
\usepackage{spr-astr-addons}
\usepackage{url}

\begin{document}

\title{2D solar modelling}
\shorttitle{2D solar modelling}
\shortauthors{Ventura et al.}

\title{2D solar modeling}


\author{P. Ventura\altaffilmark{1}}
\affil{INAF - Observatory of Rome, Via Frascati 33, 00040 Monte Porzio 
    Catone, Italy} 
\and 
\author{V. Penza\altaffilmark{2}}
\affil{Universit\`a di Roma "Tor Vergata", Via della ricerca scientifica 1,
    00133 Roma, Italy}
\and
\author{L. Li, S. Sofia, S. Basu and P. Demarque\altaffilmark{3}}
\affil{Department of Astronomy, Yale University, P.O. Box 208101, New Haven,
 CT 06520-8101, USA}

\begin{abstract}
Understanding the reasons of the cyclic variation of the total
solar irradiance is one of the most challenging targets of modern 
astrophysics. These studies prove to be essential also for a more
climatologic issue, associated to the global warming. Any attempt
to determine the solar components of this phenomenon must include
the effects of the magnetic field, whose strength and shape in the
solar interior are far from being completely known.
Modelling the presence and the effects of a magnetic field 
requires a 2D approach,
since the assumption of radial symmetry is too limiting for this topic.
We present the structure of a 2D evolution code that was purposely
designed for this scope; rotation, magnetic field and turbulence can
be taken into account. Some preliminary results are presented and
commented.
\end{abstract}

\keywords{Sun: evolution --- Sun: interior}

\section{Introduction}
It is well known that the Total Solar Irradiance (TSI) varies with
time, with variations both on a short time scale, of the order of
days to months, and on a long time scale, which follows the 11-years solar 
cycle \citep{wil01}. While the short-term variations are associated to the solar
rotation, as a consequence of surface spots and faculae, which determine
the variation of the energy flux in the Earth's direction, the reasons 
for the long-term variation are more debated, as confirmed by the extensive 
literature in that regard, ranging from a more "surface" interpretation 
\citep{krivova} to papers stressing the role of a possible variation of the 
internal structure of the Sun \citep{li01}.

In this latter case we would expect that at least the whole convective
envelope is involved, thus opening the possibility of variability on 
even longer time scales which could drive long-term variations of the 
Earth's climate on the Earth. This confirms the importance of understanding 
the role played by the internal variation of the solar structure.  

Among all the possible causes that may potentially alter periodically the
solar structure, the magnetic field is generally regarded as the main
contributor. Some preliminary investigations, limited to a 1D treatment, 
and thus limited to purely radial symmetry configurations, showed that the observed 
variation of the TSI can be reproduced by assuming the presence of a 
large-scale magnetic field, which varies in phase with the solar cycle 
(see e.g. \citet{li01}).

The 1D approach unfortunately proves to be of extremely limited relevance 
for such an important investigation, because any realistic model of the Sun, 
accounting for rotation and magnetic field, is expected to develop
asymmetries at least of the order of magnitude of the effects that we
want to investigate. Also, it is practically impossible to find any
morphology of the magnetic field that is radially symmetric, whatever
its toroidal or poloidal components are.

These arguments stimulated our group to take a step further, with the extension
of the classic YREC code for stellar evolution \citep{pierre} to a 2D structure, accounting
for variations of the main physical and chemical quantities not only
with the distance from the center, but also with the latitude.
In this paper we describe the theoretical framework that introduces 
bidimensionality, and that accounts for the effects of magnetic fields,
turbulence and rotation; we eventually provide a new set of differential
equations that must be solved to determine the solar structure, and the
changes expected when the above mentioned phenomena are considered.

The paper is organized as follows: Sections 2 and 3 review the 
modifications to the standard stellar structure equations needed to involve a 
magnetic field and turbulence; the structure of the 2D code, with the relevant
equations, is presented in Section 4. Some preliminary results are given 
in Section 5.

\section{The inclusion of magnetic field in stellar modeling}
The effects due to the presence of an internal magnetic field are
modeled following the method described in \citet{lydon}.
The magnetic field is described by means of an energy density
$\chi_m=B^2/8\pi \rho$.
We list hereafter all the modifications demanded to properly account
for the presence of this perturbation.

\begin{enumerate}

\item{
The total pressure is
$$
P_T=P_{\rm gas}+P_{\rm rad}+{B^2\over 8\pi}
$$
where the magnetic contribution was added to the standard gas and 
radiation components.
}

\item{
The equation of state (EOS) is consequently changed, as density depends
not only on pressure and temperature, but also on the strength of the
magnetic field:
$$
d \ln \rho=\alpha d \ln P_T-\delta d \ln T -\nu_m d \ln \chi_m
$$
where $\alpha=d\ln \rho/d\ln P_T$, \quad $\delta=d\ln \rho/d\ln T$, \\
$\nu_m=d\ln \rho/d\ln \chi_m$
}

\item{
The momentum equation must include a magnetic term expressing
the Lorentz force, that is 
$$
{1\over 4\pi}(\nabla \times B)\times B=-\nabla \cdot \left({B^2\over 8\pi}\right)+
{1\over 4\pi}(B\cdot \nabla)B
$$
Regarding the anisotropic pressure term \\ 
${\cal H}=\nabla \cdot (B\cdot B)/(4\pi \rho)$, the latter equation becomes:
$$
{1\over 4\pi}(\nabla \times B)\times B=-\nabla (P_m)+\rho {\cal H}
$$
}

\item{The variation of the total entropy $S_T$, in agreement with the first
law of Thermodynamics, must include not only the variation of the
internal energy and the work done by the gas to expand, but also the
variation of the magnetic energy density. Expressed differentially, the
law giving $dS_T$ becomes
$$
TdS_T=dU+PdV+d\chi_m
$$

}

\item{
The adiabatic gradient $\nabla_{\rm ad}$, defined as the logarithmic derivative 
of temperature with respect to pressure along a transformation at constant entropy, 
must change, since the entropy itself contains a new magnetic term. The complete 
expression is
$$
\nabla_{\rm ad}={\delta\over \rho c_P}\left( 1-{\nu_m\over \alpha}\nabla_{\rm m}\right)
$$
}
where $\nabla_{\rm m}=d\ln \chi_m/d \ln P$.
\item{The change in the EOS determines an alteration of the criterion for
radiative stability, because the latter is based on the evaluation of
buoyancy acting on a convective eddy. The new formulation of the 
Schwarzschild criterion is
$$
\nabla_{\rm rad}<\nabla_{\rm ad}\left(1-{\nu_m\over \alpha} \nabla_{\rm m}  \right)
$$
We note that the presence of a magnetic field with a strong gradient in the
proximity of the formal border of convection may potentially alter the
location of the convection/radiation boundary.
}

\item{The expression for the convective flux is
changed in the presence of a magnetic field. Within the context of
the Mixing Length Theory (MLT) modelling of convection \citep{vitense}, 
the convective flux $F_C=\rho v_{\rm conv}dQ$ is evaluated on the basis 
of dQ, i.e. the excess 
heat of the convective eddy with respect to the environment before dissolving: 
according to expression for the change of the total entropy given in point 4
above, inclusion of possible differences in the magnetic energy density is
required. A further modification to the calculation of the convectice flux
is needed in the evaluation of the convective velocity, made on the basis
of the work done by buoyancy: in this case the excess density is not only 
related to the excess temperature, but also on the different $\chi_m$ between 
the convective element and the environment (see point 2 above).

The above changes were used by \citet{li01} to perform a qualitative study
of the effects of the presence of a large scale magnetic field, allowed to
change periodically with the solar cycle, on the global changes of the 
solar structure. They could reproduce the observed $0.1\%$ variation of
the solar luminosity with various magnetic fields, differing in intensity
and in the region where they peak. The main results of this first investigation
was that deeper magnetic fields required higher intensity, and were
associated to stronger variation of the total radius of the Sun, than
shallower ones.
}

\end{enumerate}

\section{The role of turbulence}
From the numerical point of view, the inclusion of the effects of turbulence
is qualitatively similar to considering the presence of a magnetic field.
The key quantities in this context are the turbulent kinetic energy
$\chi_t=v^2/2$, and the degree of anisotropy, commonly expressed
via the variable $\gamma_t=1+2(v_r/v)^2$. The presence of turbulence also 
adds a new term to the pressure, i.e. $P_t=\rho (\gamma_t-1)\chi_t$. 

The modifications necessary to include turbulence are the same as those
given in the previous section, provided that $\chi_m$ and $P_m$ are replaced,
respectively, by $\chi_t$ and $P_t$, and that the variation of density in the
EOS is expressed in terms of $d\chi_t$ and $d\gamma_t$. The magnetic energy
gradient $\nabla_m$ is replaced by $\nabla_t=d \ln \chi_t/d\ln P$ in the 
evaluation of the adiabatic gradient and the super-adiabaticity, and in the 
Schwarzschild criterion.

The momentum equation is changed accordingly, the two additional terms
associated to turbulence being an isotropic pressure component and an
anisotropic term, namely
$$
-\nabla (P_t)+\rho \cal{T}
$$
where
$$
{\cal T}=\nabla \cdot [\rho (v_rv_r-v_{\theta}v_{\theta})e_{\theta}e_{\theta}+
\rho (v_rv_r-v_{\phi}v_{\phi})e_{\phi}e_{\phi}]/ \rho
$$
In the previous expression we have indicated the three components of the
turbulent velocity with $v_r$, $v_{\theta}$, and $v_{\phi}$, while 
$e_{\theta}e_{\theta}$ and $e_{\phi}e_{\phi}$ are two of the three
components of the unit tensor I.

On the basis of the numerical simulations by \citet{robin}, which give the values
of $\chi_t$ and $\gamma_t$ in the superadiabatic layer of the Sun, \citet{li02}
studied the simultaneous effects of magnetic field and turbulence, and they
presented various models that could reproduce the cyclic variation of
the solar irradiance, and a nice fit of the observed oscillation
frequencies, particularly in the 2000-4000 $\nu$Hz regime (see Fig.12
in \citet{li02}).

\section{The 2D formulation}
A 2D treatment is essential for a reliable and physical investigation 
of the effects of magnetic fields and turbulence on the solar interior,
because the 1D study does not allow for a full exploration of realistic 
magnetic field configurations, and the 2D effects may be of the same 
order of magnitude as the perturbations that we want to investigate.
In addition, the satellite PICARD, whose launch is schedule for the
end of 2009, is expected to detect other 2D observational features, 
which can also be used to discriminate among the various models.
These are the reasons behind our choice to develop a full 2D formulation.

Our approach is based on the equipotential surfaces of the gravitational
potential, $\Phi$. The mass coordinate is the mass contained within the 
surfaces $\Phi=const$, which, unlike the 1D case, may not be spherically 
symmetric. The colatitude $\theta$ is the other independent variable. 
The system is assumed to be azimuthally symmetric. Each equipotential 
surface is characterized by the expression $r=r(\Phi, \theta)$. 
The density is generally not constant on the equipotentials; yet, it 
is possible to introduce an integrated density, defined as

$$
\rho_m={1\over 2r^2}\int_0^{\pi}{\rho r^2 \sin \theta d\theta} 
$$ 

The basic equations used are the canonical relations expressing the
equilibrium of the stellar structure, namely the conservation of mass,
momentum and energy, and the Poisson equation.
Unlike the 1D case, the gradient of the gravitational potential $\Phi$
has a transverse component $\cal G$, and its radial part deviates by 
the unknown quantity U from the standard expression, $Gm/r^2$:
$$
\nabla \Phi=\left( {Gm\over r^2}+U, \cal G \right)
$$
The horizontal component of the energy flux, $F_{\theta}$, must also
be considered, although the luminosity is still related to the radial
flux.

\begin{figure*}
\vskip10pt
\includegraphics[angle=0,scale=.47]{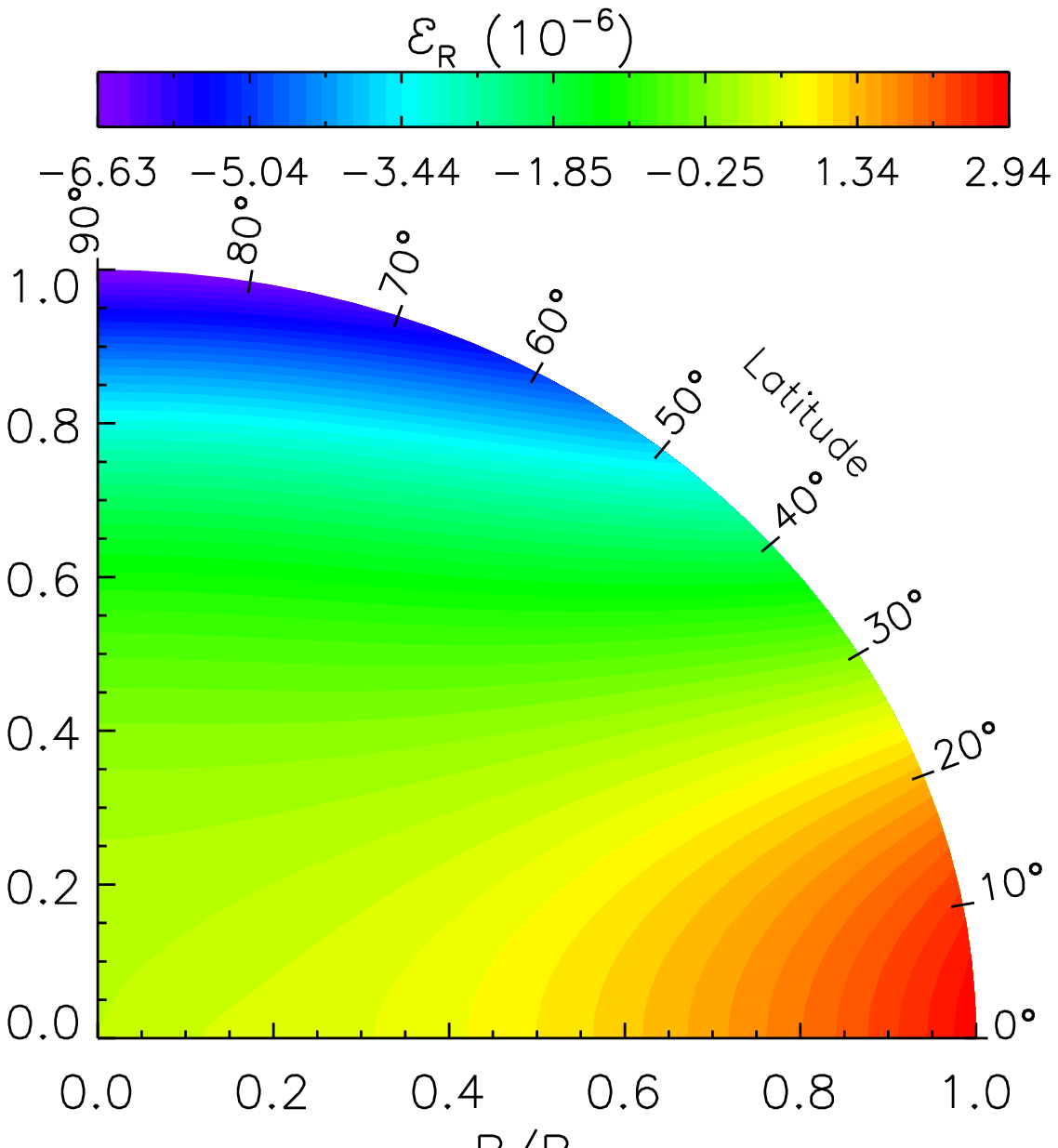}
\includegraphics[angle=0,scale=.47]{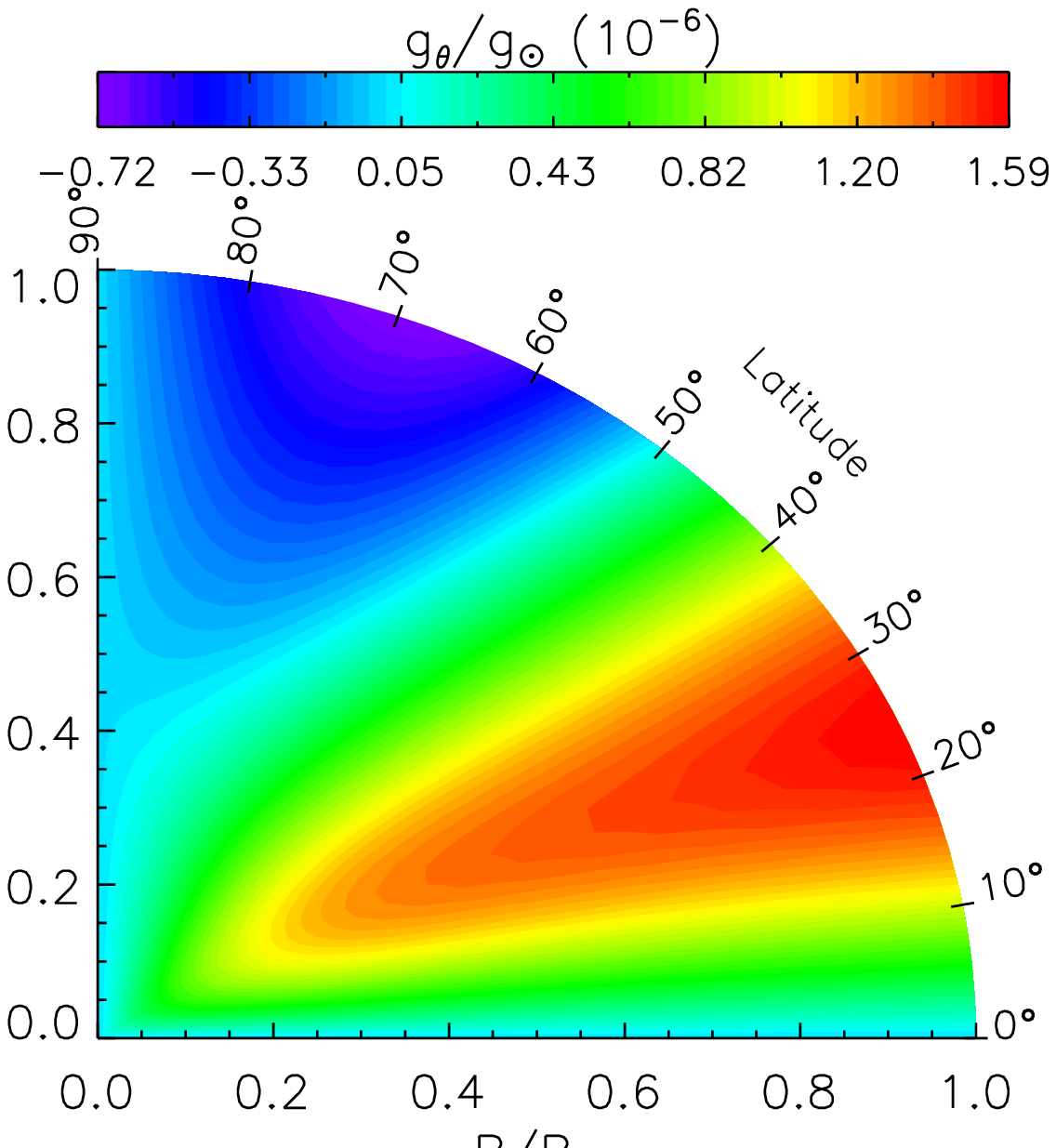}
\includegraphics[angle=0,scale=.47]{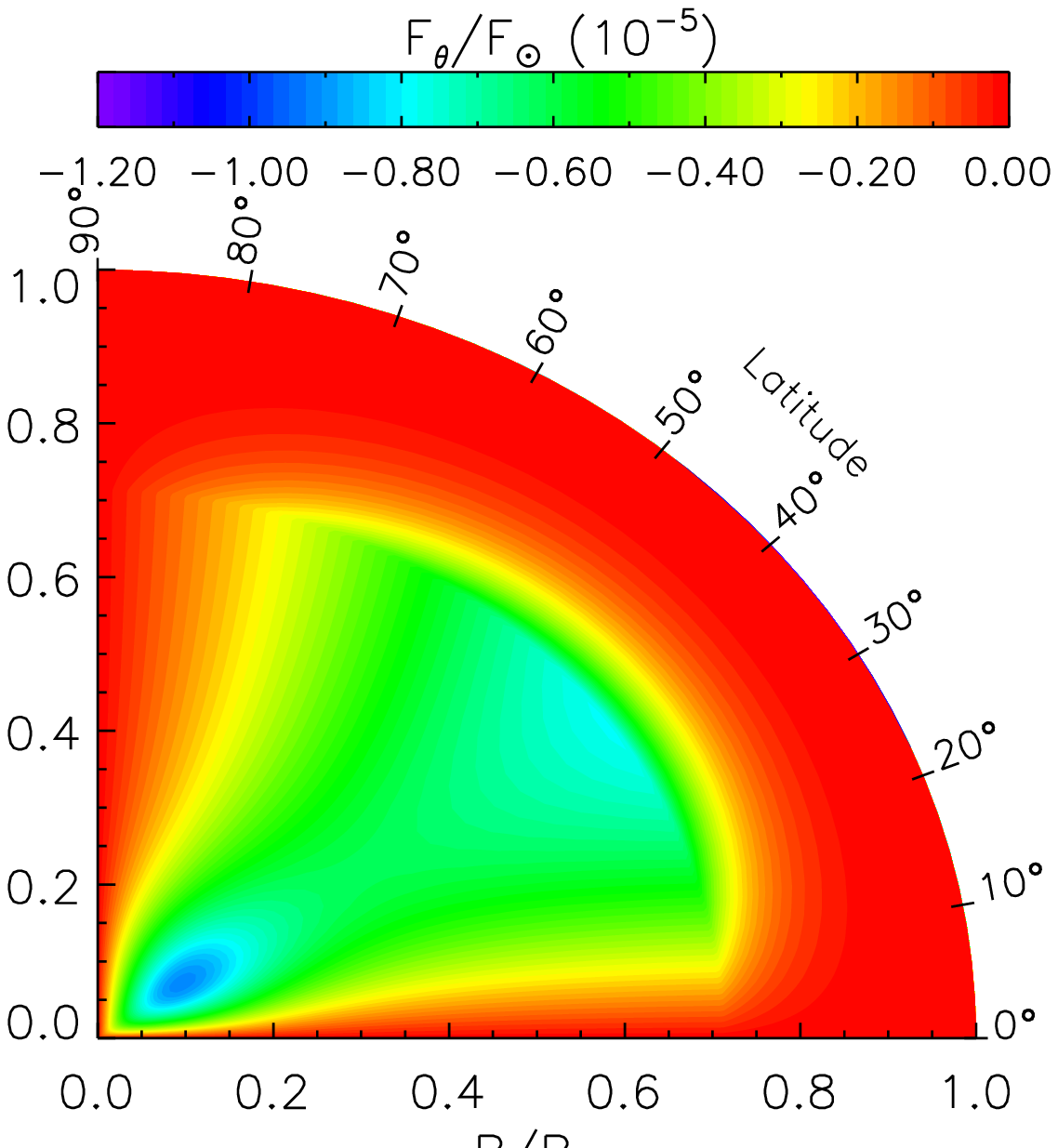}
\caption{
The structure of a rotating 2D solar model is shown in
terms of $\epsilon_R$, i.e. the variation of the radius compared 
to a non rotating 1D, radially symmetric structure (left), the 
value of the colatitudinal acceleration $g_{\theta}$ (center), 
expressed in units of the solar surface gravity $g_{\odot}$, 
and the horizontal flux $F_{\theta} (right)$.
}
\label{rotat}
\end{figure*}

In agreement with the discussion of the previous sections, the momentum
equation needs modifications with respect to the standard case to account
for the presence of magnetic field and turbulence.
Rotation generates a centrifugal acceleration that must be included:
${\cal R}=\Omega^2 r \sin \theta (\sin \theta, \cos \theta)$

Eventually, these hypotheses lead to a system of 5 differential equations:

\begin{eqnarray}
{d\ln r\over d\ln m} & = & {m\over 4\pi r^3\rho_m} \\
{d\ln P\over d\ln m} & = & -{m\over 4\pi r^2 P}{\rho\over \rho_m}({gm\over r^2}+
U-{\cal{H}}_r- \nonumber \\
& & {\cal{T}}_r-{\cal{R}}_r) \\
{d\ln P\over d\ln m} & = & \nabla {d\ln T\over d\ln m} \\
{dL\over d\ln m} & = & {m\over L_{\odot}}{\rho\over \rho_m}\left( \epsilon-
T{dS\over dt} \right)-{mF_{\theta}cot{\theta}\over {L_{\odot}r\rho_m}}- \nonumber \\
 & & {m\over {L_{\odot}r\rho_m}}{\partial F_{\theta}\over {\partial \theta}} \\
{dU\over d\ln m} & = & {Gm\over r^2} \left( {\rho\over\rho_m}-1 \right)-
{m\over {4\pi r^3\rho_m}} (2U+ \nonumber \\
 & & \cal{G}\tan{\theta}+{\partial \cal{G}\over{\partial \theta}}) 
\end{eqnarray}

The colatitudinal acceleration, $\cal G$, can be found via the horizontal
component of the momentum equation and considering the Poisson equation:

$$
{\cal G}=H_{\theta}+{\cal T}_{\theta}+{\cal R}_{\theta}-{1\over r} 
{\partial P \over \partial \theta}
\eqno(6)
$$

The 5 differential equations given above must be integrated from the center
to the surface, and from the pole to the equator. Provided that the
structure is divided into N mass shell and M angles, the linearization
procedure leads to 5(M-1)(N-1) equations for the 5MN unknown, i.e. the
values of the 5 variables in the whole star. The system is closed with
the following boundary conditions:

\begin{itemize}
\item{3(M-1) central conditions, setting R=L=U=0 at the centre.}

\item{2(M-1) boundary conditions at the surface, that express the
matching between the integration from the interior and the
atmospheric treatment.}

\item{5N conditions at the pole, i.e. that the 5 variables have
null angular gradient at $\theta=0$.}

\end{itemize}

The need of a high numerical accuracy demands one to find the
shape of the equipotential surfaces before the integration of
the above systems is performed. This is accomplished by
using the condition of hydrostatic equilibrium, based on which
we assume that pressure and its derivative with respect to mass
are constant along the $\Phi = const$ surfaces.
The interested reader may find in \citet{li06} and \citet{li09}
an accurate description of the numerical structure of the code used,
and the techniques adopted to solve the full system of linearized
equations.

\section{Applications of the 2D scheme}
The new formulation for the 2D approach was tested for two simple
cases, namely i) a solar model where only the effects of a differential
rotation in agreement with the observations of the solar spots is adopted,
and ii) a structure where a purely toroidal magnetic field is allowed
to vary in phase with the solar cycle. This step, considering the 
complexity of the numerical treatment requested, is mandatory before 
more complex magnetic field configurations, in which both the toroidal and 
the poloidal components are present, are investigated.

\begin{figure*}
\vskip10pt
\includegraphics[angle=0,scale=.70]{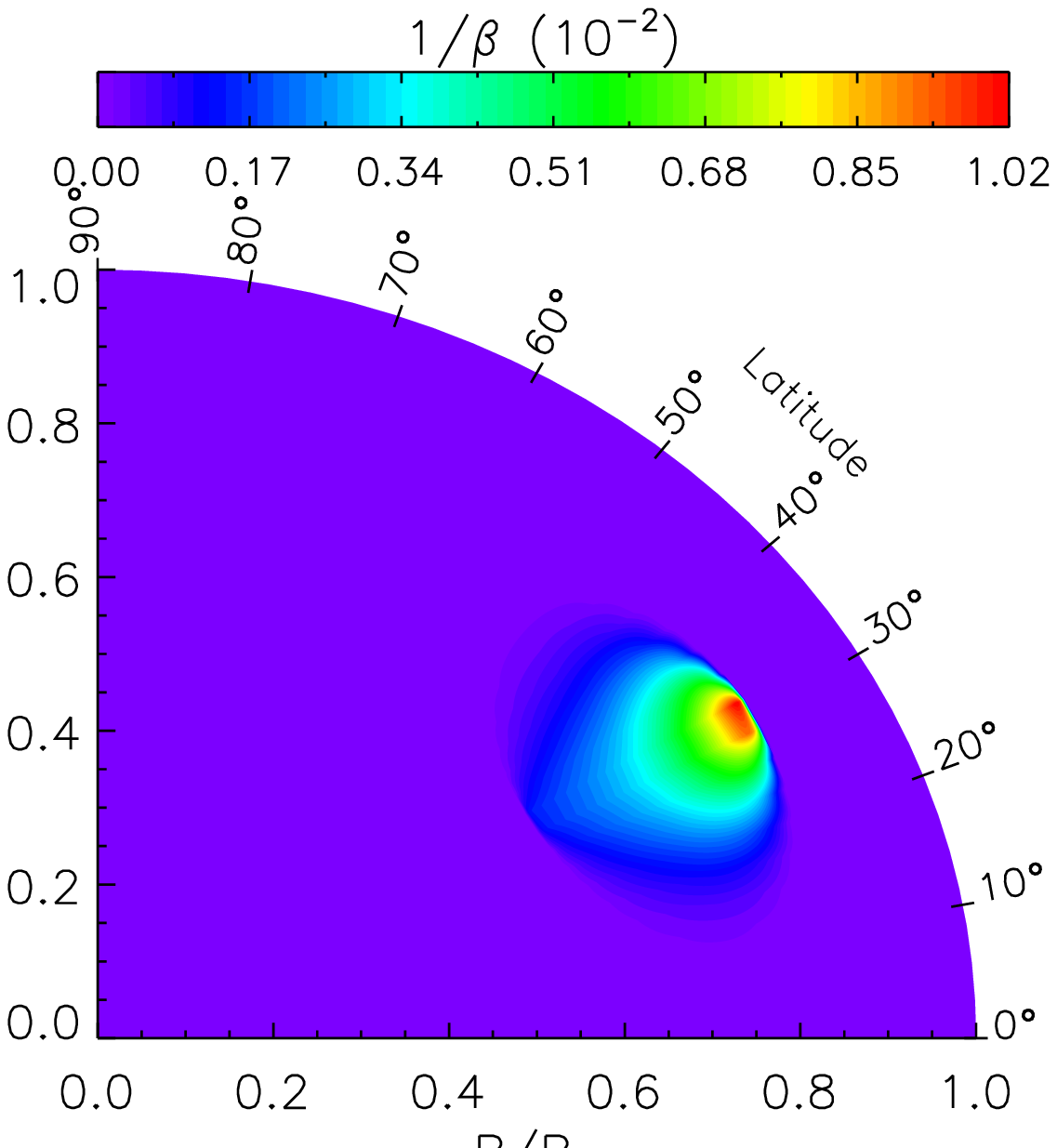}
\includegraphics[angle=0,scale=.70]{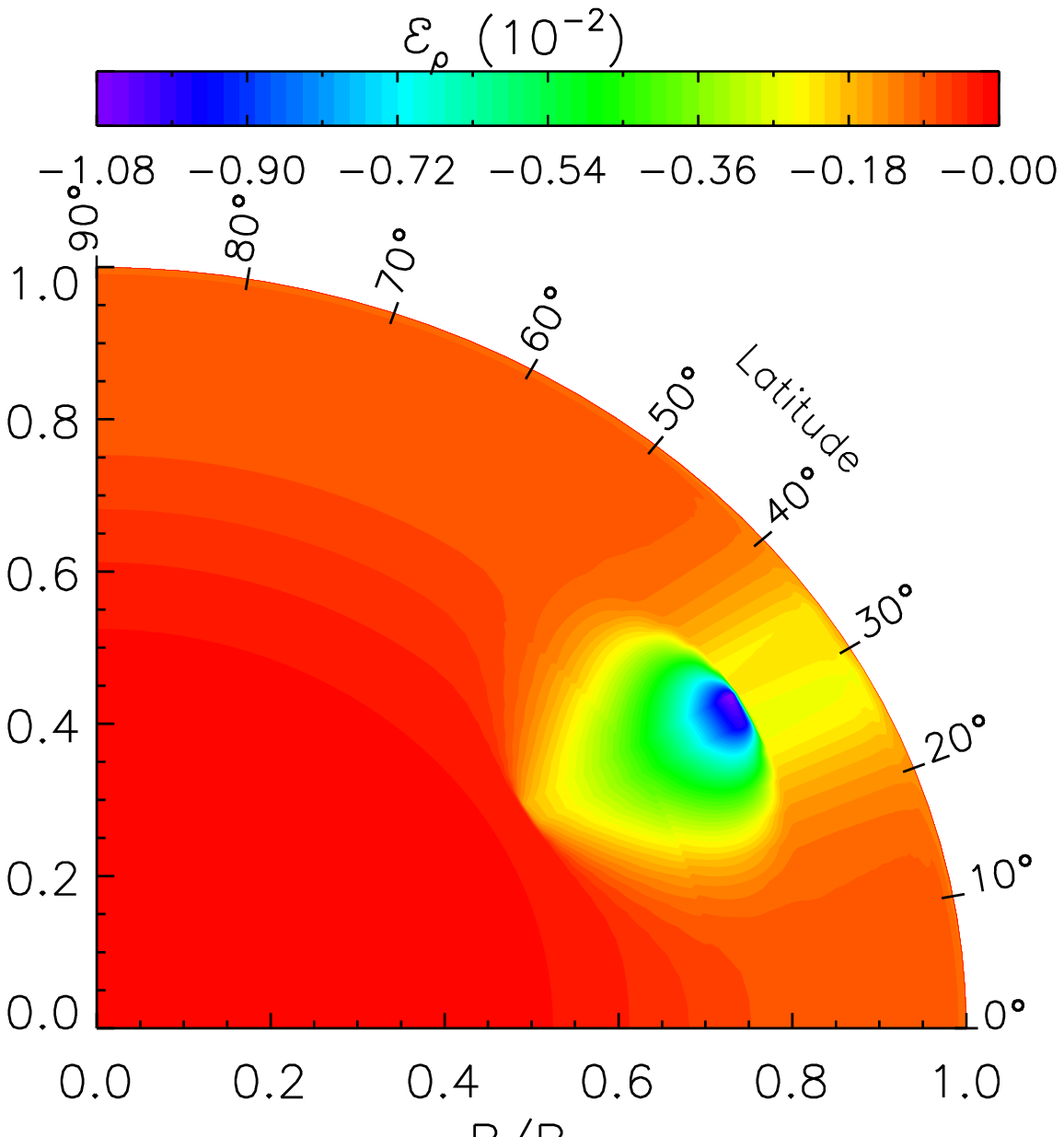}
\caption{
The structure of a solar model that includes the effects of a toroidal
magnetic field. Left: the ratio between the total pressure and the 
corresponding pressure of a standard 1D model with no magnetic field;
right: relative variation of density with respect to the standard
model
}
\label{magnet1}
\end{figure*}

\subsection{The rotating model}
We studied the effects of differential rotation on the solar structure.
No turbulence and magnetic field were included. This requires one to
neglect the $\cal H$ and $\cal T$ terms in eqs.1-5, and to consider 
only the $\cal R$ vector, associated with rotation.

The three panels of Fig.\ref{rotat} show the modifications introduced by
rotation in comparison with the standard 1D radially symmetric solar model.
We focus our attention on those quantities more sensitive to the 2D modeling,
i.e. radius (left), colatitudinal acceleration $\cal G$ (center), and
transverse radiative flux, $F_{\theta}$ (right). 

The radius variation is easily understood on the basis of the expression
for the centrifugal force, which increases from the pole to the equator, thus
determining an expansion of the star at high colatitudes, and a contraction
at the poles; this is effectively what is observed in the left panel
of Fig.\ref{rotat}.

The contours for $\cal G$, shown in the central panel of Fig.\ref{rotat},
can be interpreted on the basis of eq.6 giving the expression for $\cal G$.
Given the absence of the magnetic and turbulent contributions, the value
of $\cal G$ is given by the balance between the centrifugal horizontal term,
${1\over 2}{\Omega}^2 r \sin {2\theta}$, and the derivative of pressure
with respect to $\theta$ calculated along the constant radius surfaces.
This latter quantity is always positive, because the expansion of the 
structure towards the equator makes the $r=const$ surfaces correspond
to higher $\Phi$ surfaces (hence, higher pressure) at higher $\theta$.
The two terms are therefore of opposite sign, with the centrifugal term
dominating close to the equator, where the centrifugal acceleration is
higher. $\cal G$ is thus expected to be positive close to the equator. 
The pressure term is the dominant term close to the pole, where, indeed, 
we see from 
the central panel of Fig.\ref{rotat} that $\cal G$ is negative. Note that
$\cal G$ vanishes as the equator is approached, since we expect purely
radial contributions of all quantities there. 

The profiles of $F_{\theta}$ are associated with the pressure and temperature
profiles along the surface at constant radius. Like the pressure, even the
temperature is expected to increase moving at constant radius from the pole
to the equator; this determines a flux in the opposite direction, i.e. 
from high- to low-$\theta$ regions, which corresponds to a negative 
horizontal flux, as confirmed by the right panel of Fig.\ref{rotat}.
Again, we may comment that the effects tend to vanish at the pole and at the
equator and in the convective mantle, which is less sensitive to temperature
variation than the radiative interior.

The luminosity (not shown) changes accordingly to what is discussed above. 
We stress the increase/decrease of the radial flux on the
equator/pole, which can be directly associated to the behavior of the
radius. The nuclear core, as expected, is almost insensitive to
rotation. 

\begin{figure*}
\vskip10pt
\includegraphics[angle=0,scale=.70]{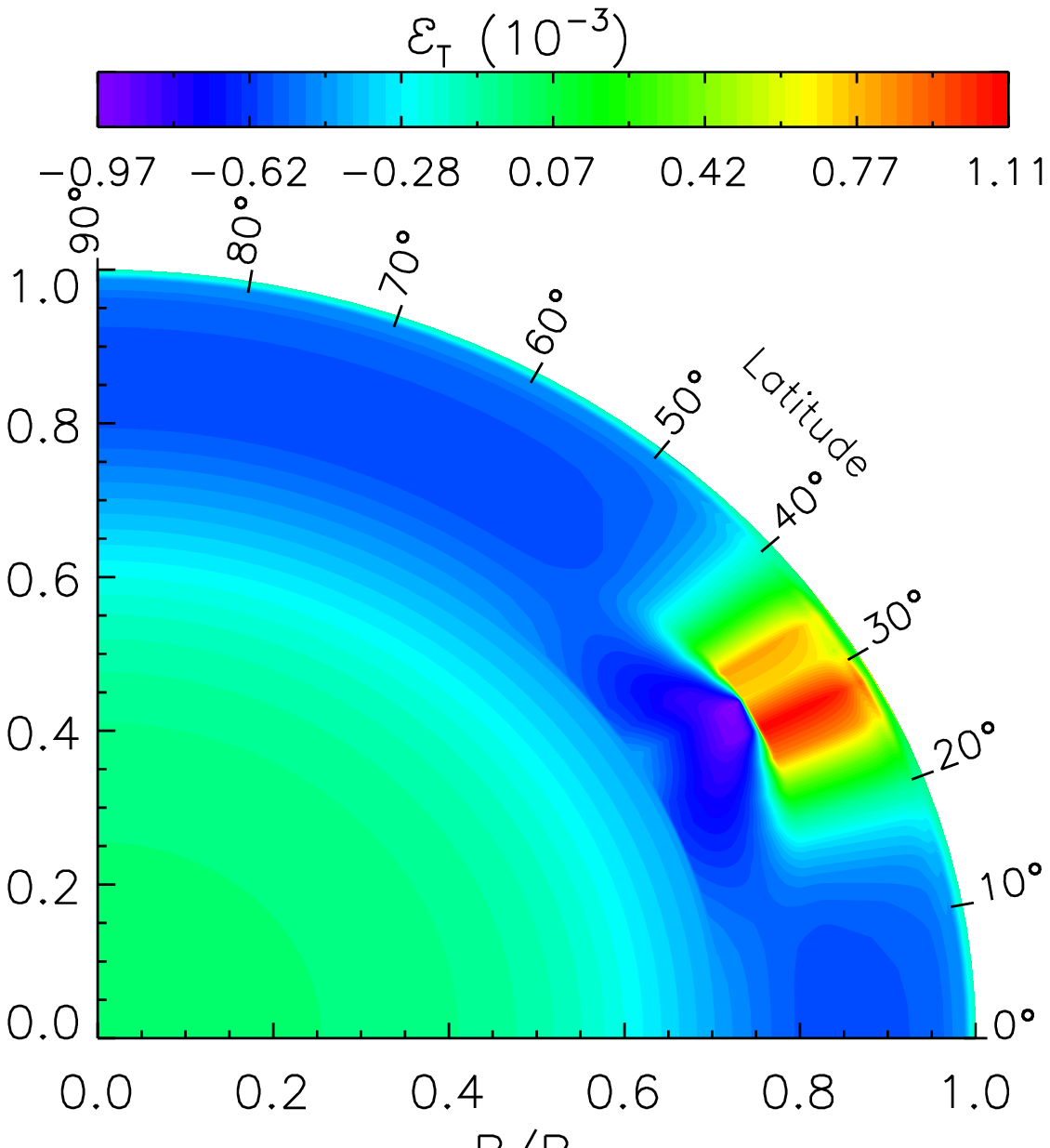}
\includegraphics[angle=0,scale=.70]{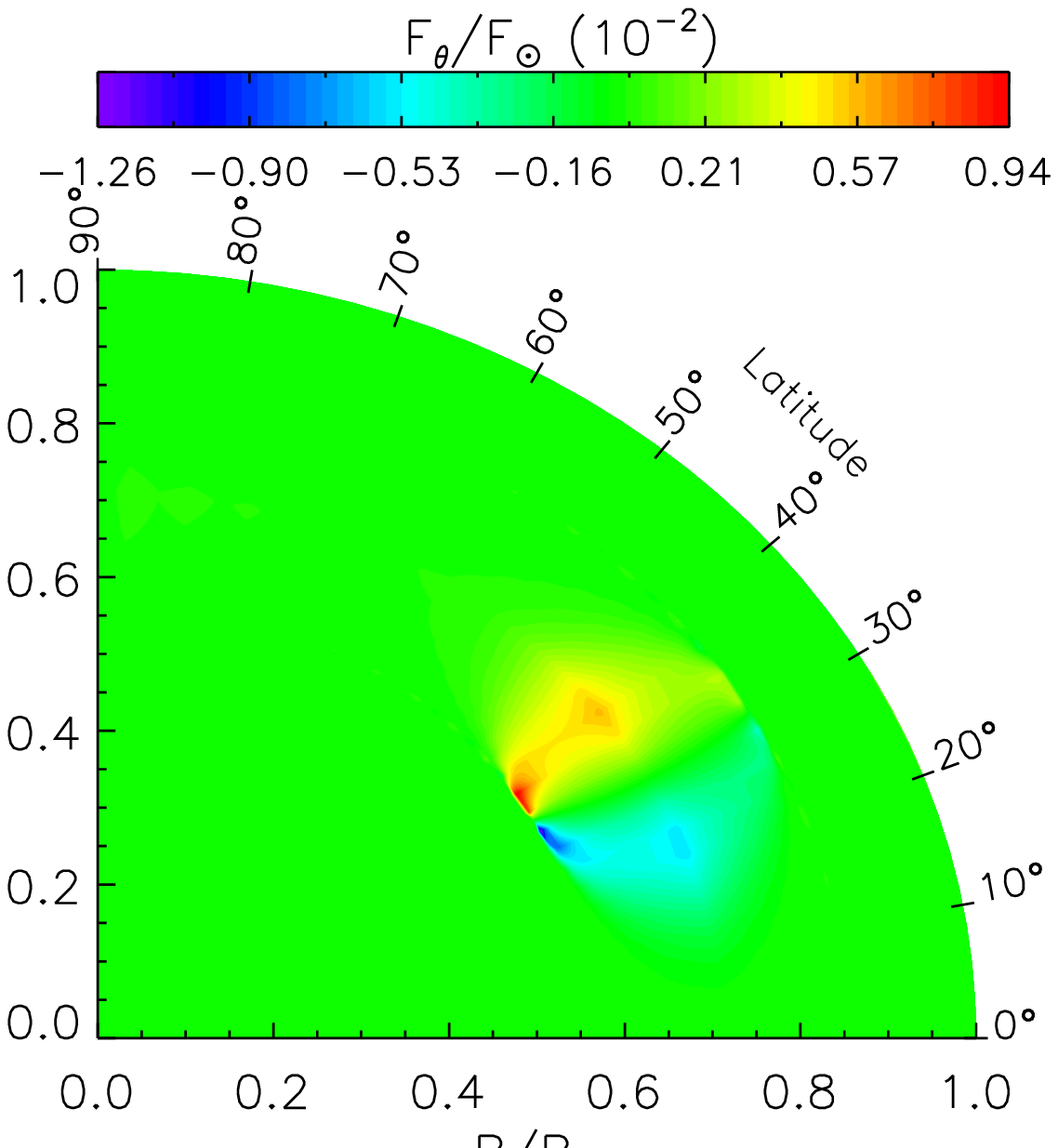}
\caption{Left: the same as Fig.\ref{magnet1}, but indicating the
relative variation of temperature; right: the horizontal component
of the radiative flux, $F_{\theta}$.}
\label{magnet2}
\end{figure*}

\subsection{Toroidal field}
We started to investigate the effects of a magnetic field with a simple
configuration, namely a toroidal field with two torus tubes that are
parallel to the equatorial plane. The field B is assumed to peak in a
region close to the bottom of the convective envelope, at a latitude
of $\theta=30^{\circ}$, and to decay gradually with mass, with a gaussian 
profile. The region where B is located can be seen in the left
panel of Fig.\ref{magnet1}, showing the ratio between the total pressure 
(including the magnetic contribution) and the pressure of the standard
1D model, in which rotation, magnetic field and turbulence are neglected.

The right panels of Fig.\ref{magnet1} shows the relative variation of
density. The density contours confirm that the region where the magnetic 
perturbation is present expands, due to the hydrostatic equilibrium condition, 
which demands a drop of the gas pressure. Note that, for continuity reasons, 
and because of the non-linearity of the stellar structure, wings of these effects
reach also zones far away from the perturbation (i.e. the polar regions
in the envelope).

The temperature (shown in the left panel of Fig.\ref{magnet2}), is consequently 
decreased in the low-density regions, although the surface zone immediately 
above the perturbation is indeed hotter, as demanded by hydrostatic equilibrium 
and the drop of density over there.
The region where the perturbation is located is thus characterized by
a strong horizontal temperature gradient, which is positive (i.e. temperature
increases with the colatitude $\theta$) in the higher $\theta$ regions,
and becomes negative from the peak of the perturbation towards the equator.
This consideration allows for a straightforward understanding of 
the right panel of Fig.\ref{magnet2}, which shows a positive transverse flux
in the upper part of the region encompassed by the magnetic field, and a
negative $F_{\theta}$ at lower latitudes.

\section{Conclusions}  
A full comprehension of the mechanism driving the long-term variation
of the total solar irradiance is far from being reached. The debate regarding
the possible role played by periodic changes of the internal structure is
still open, as confirmed by the many contributions in favor of
a completely external solution, and other works stressing the importance
to structural changes, most probably associated to a variable large-scale
magnetic field.

To understand how the presence of a magnetic perturbation could contribute
to the cyclic variation of the TSI, we developed a 2D solar evolution code,
the two independent coordinates being the mass enclosed within a given 
equipotential surface, m, and the colatitude, $\theta$. The full formulation 
consists in 5 differential equations, to be integrated from the center to the 
surface, and from the pole to the equator.

We applied this scheme to a test of the effects of a rotating solar model (note that
both uniform and differential rotation can be straightforwardly implemented),
and to a structure accounting for the presence of a toroidal magnetic field.

More general configurations of the magnetic field will be investigated
in future work. 

\acknowledgments
S.S. acknowledges the support of the G.Unger Vetlesen Foundation, and
the Brinson Foundation; L.L.H. acknowledges support by NFS grant ATM-0737770;
S.B. by NFS grants ATM-0737770 and ATM-0348837; P.D. by NASA grant NAG5-13299.


\begin{thebibliography}{}
\bibitem[Demarque et al.(2008)]{pierre} Demarque, P., Guenther, D.B. ,
Li, L.H., Mazumdar, A., \& Straka, C.W. 2008, Ap\&SS, 316, 31 
\bibitem[Krivova et al.(2003)]{krivova} Krivova, N.A., Solanki, S.K., 
Fligge, M., \& Unruh, Y.C. 2003, A\&A, 399, L1 
\bibitem[Li et al.(2002)]{li02} Li, L.H., Robinson, F.J., Demarque, P.,
Sofia, S., \& Guenther, D.B. 2002, \apj, 567, 1192 
\bibitem[Li \& Sofia(2001)]{li01} Li, L.H., \& Sofia, S. 2001, \apj, 549, 1204 
\bibitem[Li et al.(2006)]{li06} Li, L.H., Ventura, P., Basu, S., Sofia, S.,
\& Demarque, P. 2006, \apjs, 164, 215 
\bibitem[Li et al.(2009)]{li09} Li, L.H., Sofia, S., Ventura, P., Penza, V.,
Bi, S., Basu, S., \& Demarque, P. 2009, \apjs, 182, 584 
\bibitem[Lydon \& Sofia(1995)]{lydon} Lydon, T., \& Sofia, S., 1995, \apjs, 101, 357
\bibitem[Robinson et al.(2001)]{robin} Robinson, F.J., Demarque, P., Sofia, S.,
Chan, K. L., Kim, Y. C., \& Guenther, D. B. 2001, Proc. SOHO 10/GONG 2000 Workshop,
Helio- and Astroseismology at the Dawn of the Millenium, ed, A. Wilson
(ESA SP-464; Noordwijk: ESA), 443  
\bibitem[Vitense (1953)]{vitense} Vitense, E. 1953, Zs. Ap., 32, 135 
\bibitem[Willson \& Hudson(1991)]{wil01} Willson, R.C., \& Hudson, H.S. 1991, 
Nature, 351, 42 
\end{thebibliography}
\end{document}